\documentclass[12pt, draftclsnofoot,onecolumn]{IEEEtran}
\usepackage{verbatim}
\usepackage{url}
\usepackage{amsmath}
\usepackage{graphicx}
\usepackage[ruled]{algorithm2e} 
\usepackage{multirow}
\usepackage{setspace}
\usepackage{amsfonts}
\usepackage[implicit=false]{hyperref}
\usepackage{cases}
\usepackage{subfigure}
\usepackage{bm}
\usepackage{comment}
\usepackage{colortbl}
\usepackage{cite}
\usepackage{amsthm}
\definecolor{intvier}{RGB}{255,143,143}
\usepackage{booktabs}

\begin{document}
%
\title{How AI-driven Digital Twins Can Empower Mobile Networks}

\author{Tong~Li,~\IEEEmembership{Member,~IEEE,}
        Fenyu~Jiang, 
        Qiaohong Yu,
        Wenzhen~Huang,\\
        Tao Jiang,~\IEEEmembership{Fellow, IEEE,}
        Depeng Jin

\IEEEcompsocitemizethanks
{\IEEEcompsocthanksitem  T. Li, F. Jiang, Q. Yu, W. Huang and D. Jin are with Beijing National Research Center for Information Science and Technology (BNRist), Department of Electronic Engineering, Tsinghua University, Beijing 100084, China. (E-mail: tongli@mail.tsinghua.edu.cn)
\IEEEcompsocthanksitem T. Jiang is with the Research Center of 6G Mobile Communications, School of Cyber Science and Engineering, Huazhong University of Science and Technology Wuhan 430074, China. (e-mail: tao.jiang@ieee.org)
}
}

\markboth{Journal of \LaTeX\ Class Files,~Vol.~xx, No.~x, xx~2023}%
{Shell \MakeLowercase{\textit{et al.}}: Bare Demo of IEEEtran.cls for Communication Society Journals}
%



\IEEEtitleabstractindextext{%
\begin{abstract}
The growing complexity of next-generation networks exacerbates the modeling and algorithmic flaws of conventional network optimization methodology. In this paper, we propose a mobile network digital twin (MNDT) architecture for 6G networks. To address the modeling and algorithmic shortcomings, the MNDT uses a simulation-optimization structure. The feedback from the network simulation engine, which serves as validation for the optimizer's decision outcomes, is used explicitly to train artificial intelligence (AI) empowered optimizers iteratively. In practice, we develop a network digital twin prototype system leveraging data-driven technology to accurately model the behaviors of mobile network elements (e.g., mobile users and base stations), wireless environments, and network performance. An AI-powered network optimizer has been developed based on the deployed MNDT prototype system for providing reliable and optimized network configurations. The results of the experiments demonstrate that the proposed MNDT infrastructure can provide practical network optimization solutions while adapting to the more complex environment.
\end{abstract}

}

\maketitle

\IEEEdisplaynontitleabstractindextext

%
\IEEEpeerreviewmaketitle

\section{Introduction}

Mobile networks have developed rapidly over the last few decades, from First Generation (1G) voice-only networks to the current Fifth Generation (5G) wireless worldwide web networks. Following the successful worldwide commercialization of 5G in 2020, the vision and needs for next-generation mobile networks, i.e., upgraded 5G or 6G, have also been suggested. Next-generation mobile networks are expected to offer solutions for all applications~\cite{8869705}, such as networked automobiles, autonomous driving, extended reality, super-high-quality online videos, Metaverses, and a plethora of other personalized services for subscribers. In order to satisfy the aforementioned visions and objectives, next-generation mobile networks are expected to utilize extraordinarily complicated network designs that provide coexistence access to many standards and coordinate different devices. The rising complexity of networks and the diverse range of services make network planning, control, maintenance, and optimization increasingly challenging.

Conventional methodology for mobile network optimization generally follows the pathway of mathematical modeling and algorithm development. Researchers, in particular, first study the mobile network optimization problem in detail and then produce a mathematical model to capture or abstract the studied problem from physical phenomena. An optimized or approximation algorithm will be designed based on the derived model and mathematical tools. However, the extreme complexity of real mobile networks currently poses a significant threat to such a conventional methodology due to two major challenges, i.e., modeling deficit and algorithmic deficit.
\begin{enumerate}
    \item \emph{Modeling deficit.} The network optimization problems are a kind of technical problem that is hard to model, including typical issues such as network coverage, signal interference, neighboring cell selection, and handover. Current solutions mainly depend on the experience of the engineers. Next-generation mobile networks have more complex network structures and Key Performance Indicators (KPIs), making these issues more difficult. In these situations, we cannot attain an overall mathematical optimization model.
    \item \emph{Algorithmic deficit.} The majority of network optimization problems are NP-hard. As the size of the systems increases, the computational complexity required to solve this kind of problem grows exponentially. Conventional approaches reduce the computational cost of a suboptimal solution by using a static network partition. However, mobile networks are becoming increasingly complex, creating a vast space of parameters that need to be optimized. Traditional mathematical optimization tools, such as combinatorial optimization and convex optimization, are hard to apply in these situations.
\end{enumerate}

Digital twins are a promising technology that has recently caught the interest of academia and industry in mobile networks.
A digital twin is a virtual representation that dynamically updates with data from a physical twin throughout the life cycle to mimic the structure, context, and behavior of a physical object or system~\cite{9429703}. We can use real-world data to monitor, simulate, and optimize network systems' behavior by integrating digital twin technology into mobile networks and creating a network digital twin. By doing so, digital twin technology can offer new perspectives and novel approaches beyond conventional methodology to overcome modeling and algorithmic deficits.

Based on the observations made above, in this article, we propose a mobile network digital twin (MNDT) infrastructure for 6G networks. We build a network digital twin prototype system using data-driven technology to accurately model the behavior of mobile network elements (e.g., mobile users, base stations), wireless environments, and network performance. An artificial intelligence (AI) based network optimizer has been developed based on the deployed MNDT prototype system. The experimental results demonstrate that the proposed MNDT infrastructure can adapt to the increasingly complex environment of mobile network systems and provide effective network optimization solutions.

The remainder of the article is organized as follows. An overview of the proposed MNDT system is provided in the following section. The digital twin of mobile network components and the AI-based network optimizers are then specified. Following that, the prototyping system and evaluations are presented. This article is concluded in the final section.
\section{Architecture of MNDT}
\label{sec:digital}

The digital twin concept, a digital replica of one or more particular devices that can abstractly represent a real device and be used as a foundation for testing under actual or simulated conditions, was first proposed by Prof. Michael Grieves~\cite{grieves2017digital}. Developing a digital twin system usually involves three modeling aspects: a mirror model, a simulation model, and an integration model.

\begin{itemize}
    \item \textbf{Mirror model}. Building digital twins at this stage primarily concentrate on the capability of monitoring, gathering, and recording real-time behavior and the status of physical devices. More specifically, collecting historical data throughout the entire lifecycle of physical devices to create a digital mirror of those devices in virtual space.
    
    \item \textbf{Simulation model}. The digital twin at this stage is required to simulate the physical device's behavior and conduct a what-if analysis. In other words, when we apply various configurations to the digital twin, the digital twin will react and perform in a manner identical to that of the physical device.

    \item \textbf{Integration model}. At this point, the digital twin is to be an integrated model comprising a physical product and a virtual product, as well as data, services, and connections between them. Bi-directional interactions between virtual and physical products are currently the most important ones. In addition to imitating them, the digital twin must be able to control physical devices.
\end{itemize}

\begin{figure*}[tb]
\vspace{-2mm}
\centering
\includegraphics[width=0.9\textwidth]{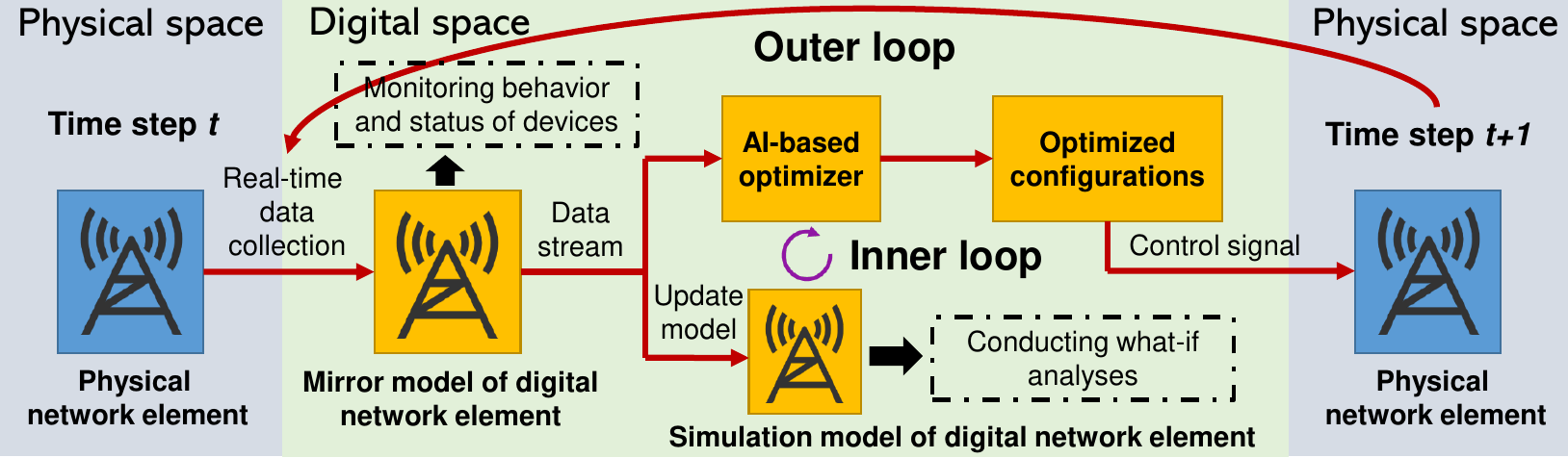}
\caption{An integration model architecture of Mobile Network Digital Twins (MNDT).}
\label{fig:MNDT}
\vspace{-2mm}
\end{figure*}

In this paper, we propose an integration model architecture of mobile network digital twins. The architecture is shown in \figurename~\ref{fig:MNDT}, having a two-loop structure: an inner loop and an outer loop. After monitoring the behavior and status of physical devices, mirror models of digital network elements are first updated in the virtual space at time step $t$ based on real-time data from physical network elements. After that, the inner loop component will receive the data stream. The inner loop, i.e., the simulation-optimization module, consists of an AI-based optimizer and a simulation model of digital network elements. Specifically, what-if analyses are done using a simulation model. AI-based optimizers work iteratively with the simulation model to find the best configurations for mobile networks. Digital twins then put the configurations into action to manage the physical devices. On the other hand, the outer loop part is for updating mirror models and simulation models of digital twins based on the feedback of the real-world network performance. In particular, based on the real-world network data at time $t+1$, the network state is collected, and the mirror model is updated accordingly. The difference between the performance of a real-world network and the performance predicted by simulation models is evaluated. Based on the differences, the inner loop module is then optimized in a series of steps to achieve integration between physical and virtual space.

\begin{figure*}[tb]
\vspace{-2mm}
\centering
\includegraphics[width=0.7\textwidth]{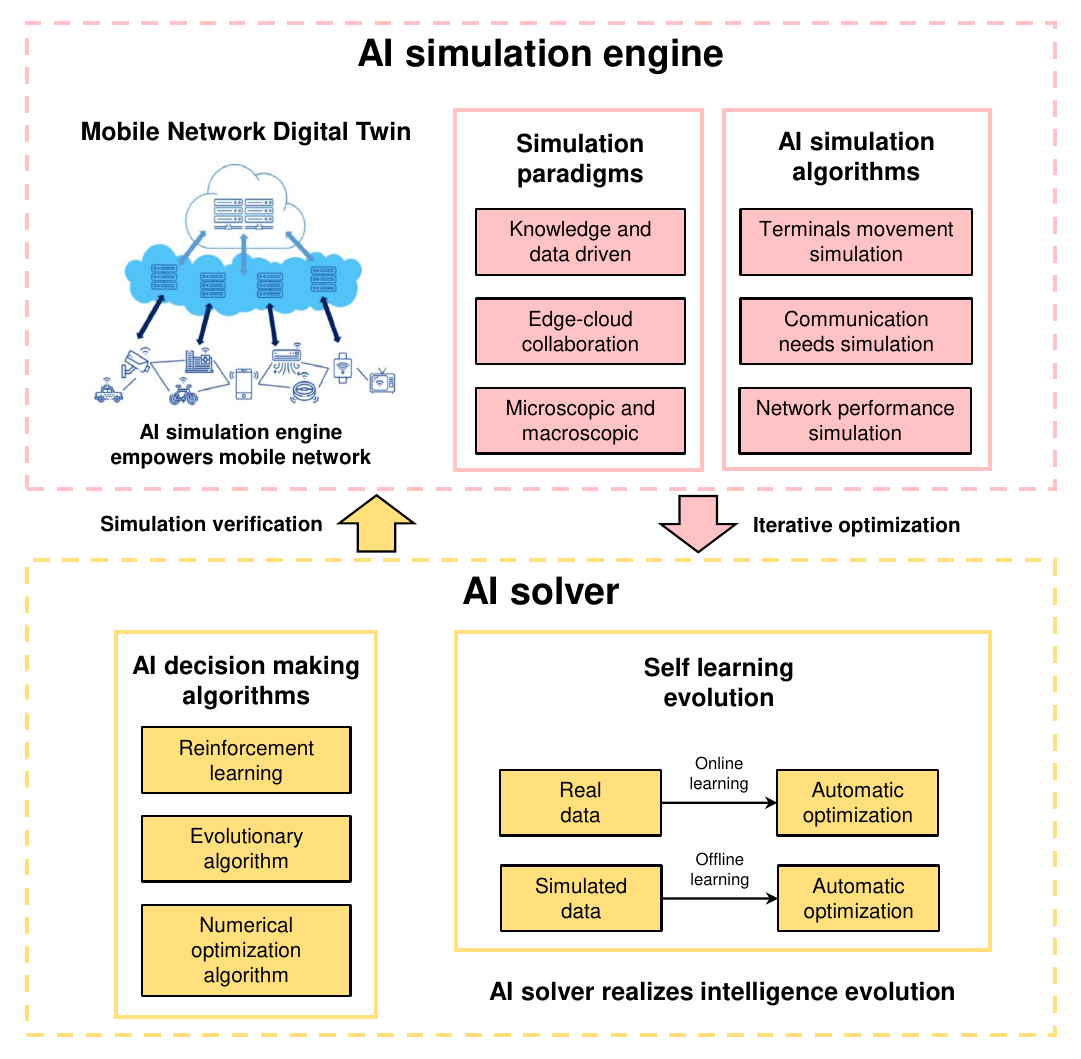}
\caption{The simulation-optimization module of mobile network digital twin.}
\label{fig:simulation_optimization_module}
\vspace{-2mm}
\end{figure*}

We next proceed to detail the inner loop, i.e., the simulation-optimization module, which is the core component of MNDT to address modeling and algorithmic deficiencies in mobile network optimization. As shown in \figurename~\ref{fig:simulation_optimization_module}, the simulation models of digital twins and AI-based optimizers serve as crucial building blocks for network intelligence. 

The ability of network devices to sense the state and collect data has significantly improved as a result of the promotion and growth of smart terminals, IoT devices, and sensors in recent years. This improvement paves the way for realizing the mirror model of network digital twins. However, relying solely on the real-world data of networks does not effectively and accurately reflect the operation dynamics of the mobile networks' entire life cycle. Some what-if operations, such as changing network configuration and updating network topology, are practically impossible to obtain due to the limitations of data acquisition capability and data privacy concerns. As a result, generation technology has also grown to be a crucial component of the network digital twin simulation model implementation. With generative techniques, we can perform a counterfactual situation extrapolation and generate data similar to the real situation by modeling the movement of terminals in the network, the communication needs of people, and the performance of the network. By doing this, we can conduct what-if analysis using simulation models to verify or evaluate the optimized configuration offered by the AI-based optimizer without mathematically modeling the mobile networks.

AI-based optimizer is responsible for analyzing and processing the simulation results, considering the simulation process and the realistic requirements to provide the best solution to the optimization problem. Specifically, reinforcement learning algorithms are frequently used in many fields to solve decision problems thanks to the recent rapid development of AI technology. The advancements and significant changes made to computer vision, natural language processing, and other fields demonstrate the potential and worth of these algorithmic technologies. In summary, the simulation engine offers simulation validation for the decision outcomes of the AI-based optimizer, and the AI-based optimizer iteratively optimizes based on the simulation engine's feedback.

\section{Digital Twin of Mobile Network Elements}
\label{sec:temporal}
Due to the simplified simulation of wireless communication procedures and the inconvenient deployment of optimization algorithms on the current network, the conventional network simulation platforms perform with low fidelity and cannot perform the evaluation results in real time. To overcome these drawbacks, we adopt the concept of a digital twin, which has been utilized in aerospace, urban construction, and industrial plant operations, establishing the mirroring of physical mobile network entities in the virtual space. Via collecting mobile network operational data and understanding the underlying mechanisms, we decompose the mobile network into three types of elements: the receivers (mobile users), the transmitters (base stations), and the communication channel (wireless environments). Notably, we consider the downlink transmission in our case. We set up a virtualized version of each element by modeling its principle and configuring it with empirical data. By assembling these virtual elements, we form the overall MNDT system. In this section, we elaborate on establishing the digital twin for each element.

\subsection{Digital Twin of Mobile Users}
As an essential component of mobile networks, users are characterized by various complicated properties, such as moving trajectories and communication traffic demand over a period of time. However, these data are generally unavailable because of the privacy issues in data sharing and the high expenses in data acquisition, resulting in the generative moving trajectories and communication traffic demand as alternatives.


To model their moving behaviors, we classify the movement of mobile users into two categories: outdoor (on the road) and indoor (within an AoI). Regarding the outdoor simulation, according to the generated travel schedule and movement type based on~\cite{yuan2022activity}, we simulate the mobile trajectories with details. Specifically, following the shortest path calculated by A* algorithm~\cite{hart1968formal} between the origin-destination (OD) schedule, we obtain the route. For the pedestrians or vehicles on the route, their positions at an arbitrary time are determined by the Krauss model~\cite{krauss1997metastable}, widely leveraged in vehicle simulation. On the other hand, for the indoor simulation, the repulsive interactions between mobile users and obstacles and the attractive interactions between users and their destinations are considered. Jointly integrating the neural networks and discovered physics, we can achieve both high generalization and fidelity~\cite{zhang2022physics}.


Analyzing the characteristics of real-world communication traffic demand, especially in time series, we find that the traffic demand exhibits a mixture of multiple behavioral patterns, while different mobile users may share similar traffic patterns. Accordingly, we implement a multi-scale hierarchical structure for users' demand generation, including multiple pattern generators and switch mode generators on the microscopic level and traffic demand patterns clustering on the macroscopic level. Consequently, we can generate large-scale mobile users' traffic demand with variety and fidelity.


On the data generated for a specific mobile user, we can estimate the communication rate if it is connected to a base station with given operating parameters. Moreover, to ensure the quality of service, the estimated transmission rate should meet the user's traffic demand.


\subsection{Digital Twin of Base Stations}
Base stations perform an important role within mobile networks, supporting the uplink and downlink of wireless communications with mobile users by multiple access technologies. To build the digital twin of base stations, we aim to equip them with proper functions and configure them with realistic settings. We set its GPS coordinates for a digital twin base station based on a real base station location. And other configurable engineering parameters such as transmit frequency, transmit power, antenna height, antenna azimuth, antenna elevation, and antenna pattern are set with default values according to the real data collected by the operators. We implement the functions of multiple access, user association scheduling, user mobility management, etc., as well. The simulation of a wireless transmission initiated by a user who wants to access online services through cellular networks. On recognizing the service type, the digital twin base station adjusts its transmit power and frequency bandwidth to meet the requirement of transmission rate to satisfy the user's quality of service. The adjustments are determined by the Shannon Equation, which indicates the maximum allowed error-free transmission rate in wireless communications. When plenty of users are accessing services simultaneously within a specific region, it is challenging to sophisticatedly allocate the limited resources in order to achieve the desired transmission rate for each user, thus requiring the design of corresponding optimization algorithms. Basically, in terms of user association, we implement the nearest base station and random sub-carrier access approaches as fundamentals. Moreover, if no optimized method is involved, the base station adopts a fair transmit power allocation strategy regarding its connected users.

Once a specific base station's engineering parameters are determined, we can derive its wireless signal strength coverage area with the help of channel state information. During the transmissions between multiple users and multiple base stations, we monitor the statistics like the summation of base station rate, transmit power consumption, user rate, overall network throughput, and user outage ratio. By observing this information, we can not only surveil the MNDT state but also evaluate the performance of any proposed optimization algorithms. In our proposed MNDT system, we divide the whole procedure into time steps at the level of seconds. Within each time frame, it is feasible to deploy the algorithms by re-configuring the base station engineering parameters. Following this way, the solutions for a variety of optimization problems, such as resource allocation, offloading, energy-efficient transmission, interference alleviation, and coverage enhancement can be tested and evolved quickly, bridging the gap between virtual digital twin space and the mobile network reality with a seamless connection.

\subsection{Digital Twin of Wireless Environments}
A high-resolution and accurate description of the mobile network benefits the system design and performance optimization of wireless communications. Given the user location, base station location, related engineering parameters, and scenario, especially buildings layout, the dominant factor affecting the communication quality is the channel status. We focus on modeling the channel, in particular, on investigating how the transmit signals attenuate within the real environment. The transmitter produces a planar electromagnetic wave for a wireless link, and then the carrier frequency modulates it before being sent out. On the way traversing towards the receiver at nearly light speed, the wireless signal interacts with any obstacles it encounters. Finally, the signal is induced by the receiver's antenna and then demodulated for further processing. As a result of reflection, refraction, and diffraction, the signal strength attenuates. And consequently, generate secondary scattering signals. Whether there exists a Line-of-Sight (LoS) transmit route or not is essential to the received signal strength level.

The overall path loss in dB is
\begin{equation} \label{equ: path loss}
PL = L_d + L_s + L_f(t),
\end{equation}
where $L_d$ is the path loss in free space, $L_s$ is the shadowing attenuation, which is caused by large-scale objects like buildings or mountains. $L_d$ and $L_s$ jointly contribute to the large-scale fading. Furthermore, $L_f(t)$ is the small-scale, time- and frequency-dependent fading caused by small transient objects. Generally, it is unfeasible to describe $L_f(t)$ explicitly; therefore, Rayleigh, Rician, and m-Nakagami distributions are considered instead.

Channel modeling is commonly classified into two categories: the stochastic model and the deterministic model. They both estimate the large-scale fading but concentrate on different modeling scales. The stochastic model uses empirical equations to describe the path loss from the macroscopic view. Though the calculation is efficient, the models are designed for general cases hence suffering from low fidelity. The Finite Difference Time Domain (FDTD) is a representative approach of the deterministic model. It rasterizes the region of interest into small grids, and within each grid, it utilizes the discrete approximation form of the Maxwell Equations, along with the boundary conditions, to infer the electric and magnetic fields over the spatial and temporal dimensions. The FDTD method models the channel from the microscopic view. However, FDTD is both time and computationally memory-consuming, limiting its applicability in a larger scenario. Raytracing is another deterministic model which lies between the microscopic and macroscopic scales. It traces the multiple signal rays, including LoS, reflection, diffraction, and scattering ones, traveling from the transmitter to the receiver. The channel state can be derived by calculating the overlapped effect of these rays.

In order to construct the digital twin wireless environments, we take account of the real scenario information downloaded from Open Street Map, together with the raytracing technology, to model the path loss with the digital twin base stations configured.

\section{AI-Based Network Optimizer}
\label{sec:RL}

Through simplification and approximation of the network system, many decision-making tasks in this scenario are transformed into numerical optimization problems, and then decisions are made by solving the numerical optimization problems. However, the computational cost of traditional numerical optimization methods often increases geometrically with the scale of the problem, which means that such solutions are difficult to solve large-scale problems in real-time. In addition, the application premise of this type of scheme is that the form of the optimization objective is very clear and can be simplified and approximated to a relatively simple numerical optimization problem. These conditions limit the use scenarios of this type of scheme.

Compared with traditional numerical optimization methods, deep reinforcement learning has no specific requirements for the form of optimization objectives, and deep networks trained with rich and diverse samples often have certain generalization capabilities and can be directly applied to new models without additional training. The inference process of the deep network is relatively fast, so it can quickly give a resource allocation plan for large-scale problems.

\subsection{Reinforcement Learning Based Dynamical Wireless Network Resource Allocation}
Dynamical Wireless Network Resource Allocation aims to optimize user association, resource block selection, and power allocation at each timestep to improve the user service quality.
Consider there exist numerous moving users, each with a specific flow demand and over a hundred base stations, each with several resource blocks in a certain area.

We build the above resource allocation problem as a Constrained Markov Decision Process~(CMDP), which is composed of state space, action space, reward function, and cost function.
State space means all possible states of the system. The state consists of remaining time, the remaining demands of users, the maximum transmit powers of all base stations, and the signal decay factors from each base station to each user.
Action space means all possible decisions. The action is a vector consisting of user association, resource block selection, and power allocation.
The reward function indicates what kind of reward signal the system will provide feedback after receiving different input actions. The reward function is equal to the total throughput of the users.
Like the reward function, the cost function indicates what kind of cost signal the system would receive.
The difference between these two functions is that the cumulative rewards are required to be maximized, while we should limit the cumulative costs within a specific range.
In this problem, the cost function is the sum of each user's satisfied demand, which is equal to the minimum unmet demand and the user's throughput, and the cumulative costs should be more than the total user demands.
The transition function denotes how the system's state would change with input actions: the signal decay factors change with the users' locations, and the remaining time minuses one.
Moreover, there are multiple extra constraints in this allocation problem: a user can only be connected to one base station, and a base station can only access a specific amount of users at most; each resource block under the base station can only be allocated to one user accessing the base station, and each user can only select one resource block; the sum of the transmit powers allocated by users accessing a base station cannot exceed the maximum transmit power of the base station.

Numerous and diverse constraints make the problem difficult to solve, so we consider decoupling them and optimizing the separated problems.
The constraints for the cost function are multiple-step constraints, and the constraints on actions are single-step ones.
We introduce a new variable, single-step demand, to decouple the multiple-step and single-step constraints. At each timestep, the policy generates a single-step demand for each user and then generates the action motioned above based on the single-step demand.
The single-step demands should be optimized to maximize the total throughput of the users in an episode while ensuring that the original multiple-step constraints are satisfied.
It is a constrained MDP; we solve it with a safe RL method, like P3O~\cite{shen2022penalized}.

As the dependencies between the tasks of user association, resource block selection, and power allocation, the agent decides them sequentially under the constraint of single-step demand.
For the tasks of user association (or resource block selection), the single-step demand constraint and the constraint of discrete actions exist.
We utilize a neural network to evaluate each user's reward and cost for selecting different actions.
The users and the base stations (or the resource blocks) can be considered as two node sets, and the edges only exist between any two nodes from different sets.
The task can be seen as a bipartite graph matching problem if we set the weights of the edges as the weighted sum of the predicted reward and cost.
We solve the matching problem using the Hungarian algorithm and then decide for each user which base station (or resource block) to select according to the solution.
For the task of power allocation, we build a policy network to decide the power allocated to each user and train it with an RL method.
The RL method is similar to DDPG~\cite{lillicrap2015continuous}, but the action-value network is replaced with an explicit computational form.

\subsection{Reinforcement Learning Based Sleep Mechanism}

\begin{figure*}[t]
	\centering
	\subfigure[Prototype system.]{\label{fig:prototype}\includegraphics[width=.65\textwidth]{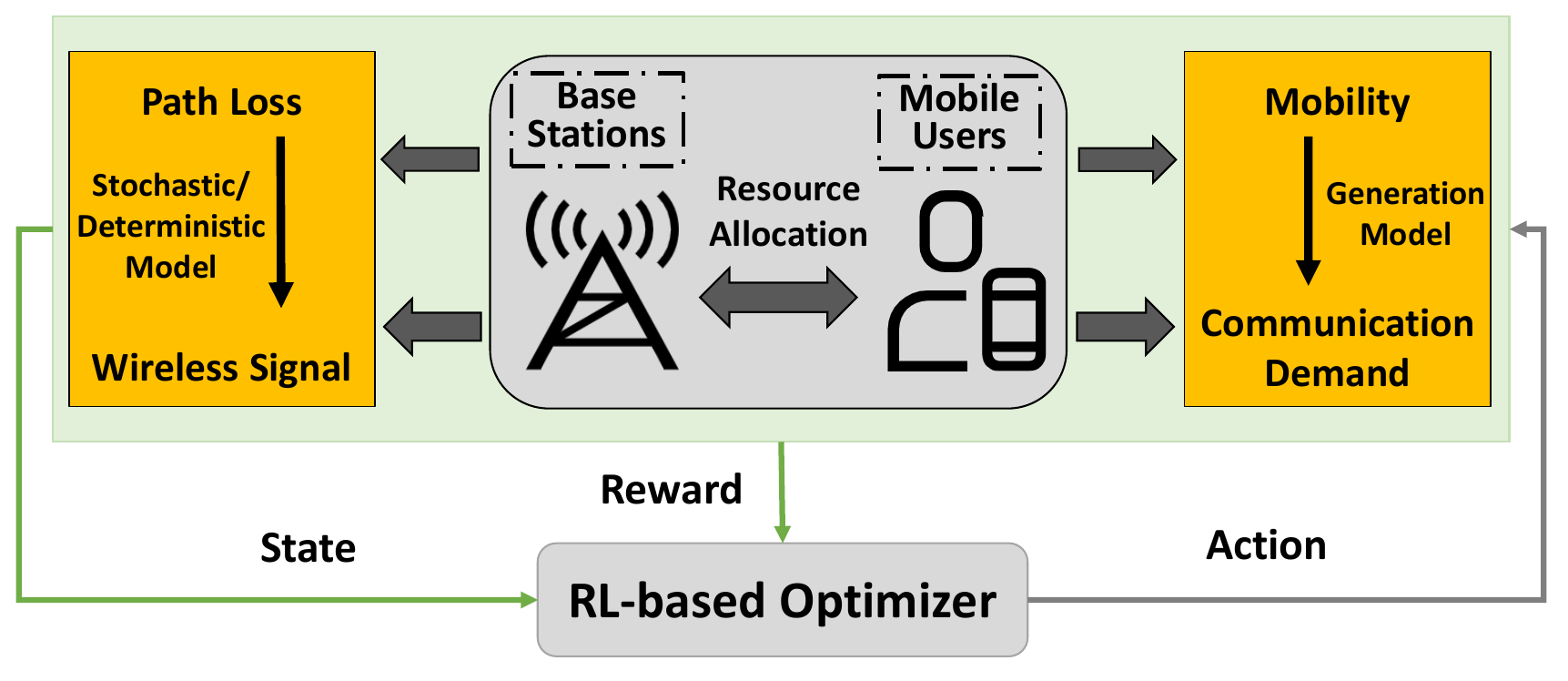}}
   \subfigure[Interaction process.]{\label{fig:interaction} \includegraphics[width=.9\textwidth]{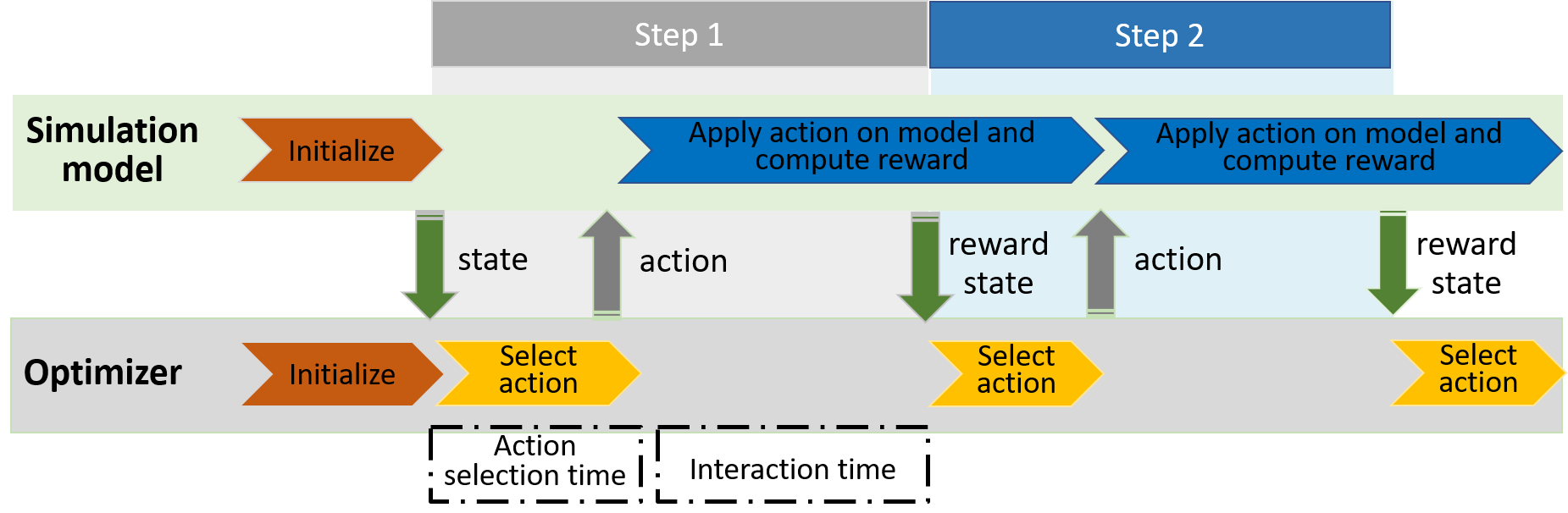}}
\caption{The architecture of prototype system and interaction process between simulation engines and AI-based network optimizer.}
\label{fig:gym}
\end{figure*}

The base station is one of the core parts of the mobile network, and it continues to increase as the network demand increases. However, the total demand of users generally changes periodically with time and shutting down the base station and internal components, like cells, can effectively save energy consumption.

The method in this subsection divides all cells into multiple grids and then trains RL agents to decide whether each cell should be opening, sleeping, or closing at each time step.
If all cells in a base station were off, we would set the base station to off. Otherwise, we would set it to on.
The total traffic load in each grid would be assigned to the cells proportional to their capacities.

We consider the above control task of the cells as an MDP.
In this problem, state space, action space, reward function, and the cost function is defined as follows.
The state consists of the timestep, the last traffic demand in all grids, the device parameters of the entire cells, and the control status of all cells and base stations.
The action is a vector that sets the cells to different control statuses. Total available control statuses are `on', `sleeping', and `off'.
The decision's first goal is to save the entire network's total energy, including the RRUs, BBUs, and air conditioners.
The second goal is to avoid frequent switching of the states of the base stations and cells.
Thus, the reward function comprises the negative value of the total power
and the negative value of the switch costs of cells and base stations.
The former relates to the action, while the latter relates to the difference between the last and current actions.
The transition function is relatively simple—the traffic demand changes according to the collected data in Nanchang. The control status of cells is changed according to the current action. Base stations switch on or off depending on the control status of the cells in each of them.

The number of cells is too large for this problem, so it is infeasible to train an agent to control all cells directly.
So we build an agent for each cell, and these agents share the same policy and action value networks and experiences.
The decision of each agent can only directly control the state of the base station to which it belongs. Thus, the reward for each agent is set to the sum of the power cost itself and the base station.
The input of the policy network is the information according to a single cell, including timestep, the traffic demands of the grid in the last four steps, the cell's device parameters, and the cell's control status and the base station to which the cell belong.
The input of the action value network extra included two masks.
The first mask means whether the traffic demand of the grid could be satisfied if the cell is sleep or closed.
The second mask means whether the base station could be closed if the cell is closed.
These two masks cover the impact of the control status of other cells on the cell's reward.
The policy and action value networks are optimized similarly to Mean Field RL~(MF-RL)~\cite{yang2018mean}. 
Compared with the mean action proposed in MF-RL, the two masks mentioned above can better characterize the overall impact of decisions made by other agents.
\section{Prototype System and Evaluations}
\label{sec:sys}

\subsection{System Deployment}

To verify the efficacy and efficiency of MNDT, we develop a prototype system featuring the complex interactions between mobile users and base stations, meanwhile monitoring their status.

As illustrated in Figure~\ref{fig:prototype}, we treat the procedure for resource allocation as a three-phase process: i) For an arbitrary mobile user, it accesses a base station based on a sorted order; ii) Determine whether the selected base station is eligible for the wireless transmission by checking the allocated transmit power and resource blocks; iii) After checking all the possible base stations, the resource allocation plan for the mobile user is updated. A mobile user is associated with a base station and resource blocks through multi-step iterations. As mentioned in Section~\ref{sec:digital}, the base stations, the mobile users, and their attributes, together with the integrated AI-based network optimizer, constitute the whole prototype system. During the optimization process, the mobile network updates the states and rewards according to the last time step action, then feeds them to the network optimizer to choose the next step action.

Specifically, within each episode, there is a fixed number of steps in the training process, and each step consumes the same simulation time, as shown in steps 1 and 2 in Figure~\ref{fig:interaction}. During each step, the states and rewards are transferred to the optimizer at the beginning as inputs to the neural network and obtain the resulting actions for computing the subsequent step rewards by the simulation engine. The action selection time and the interaction time indicate the duration of selecting actions and the remaining time within each step, respectively.


For ease of demonstration, we choose a relatively small area ($2\times2~{\rm km^2}$) and the mobile users, lanes, AOIs, and base stations (145 indoor and 39 outdoor) within it, as displayed in Table~\ref{tab:settings}. Meanwhile, some other parameters related to wireless transmission are covered as well.


\begin{table}[t]
\begin{center}
\caption{System settings}\label{tab:settings}
\begin{tabular}{|c|c|c|c|}
\hline
Area ($\rm km^2$) & $2\times2$ & Number of users & 13,000   \\\hline
Number of lanes & 1,850 & Number of AoIs & 1,417    \\\hline
Number of base stations & 184 & Channel number & 45  \\\hline
Noise density (dBm/Hz) & -174 & Bandwidth (Hz) & 1.8e5 \\\hline
Max Transmit Power (dBm) & \multicolumn{3}{c|}{Outdoor: 30, Indoor: 24}  \\\hline
\end{tabular}
\end{center}
\end{table}

\subsection{Experimental Analysis}

In order to verify the effectiveness of the proposed RL algorithm, we will briefly introduce the setting of the problems of Dynamical Wireless Network Resource Allocation and Sleep Mechanism and then demonstrate the performance of heuristics and RL algorithms.

\paragraph{Dynamical Wireless Network Resource Allocation}

Consider there are $8000$ mobile users, each of which has a flow demand, and $184$ base stations, each of which has $184$ resource blocks in an area of $2km\times2km$.
We assume that the users move in a straight line at a uniform speed in the area and randomly generate each user's starting and ending points.
The total demand of each user in the $20$ steps is uniformly sampled from 0 to 60*Bandwidth.
The agent trained with our method should allocate the suitable resource to users so that the satisfaction ratio is not lower than 95\%.

We compare our method with two baseline methods.
The former method equally divides total demand for each step, while the latter only maximizes the total throughput and ignores users' demands.
The experiment result is shown in table~\ref{tab:allocation_results}.
The result demonstrates that our approach maximizes the total throughput while satisfying the constraints.

\begin{table}[t]
\begin{center}
\caption{The setting of Dynamical Wireless Network Resource Allocation}\label{tab:allocation_results}
\begin{tabular}{c|c|c}
\toprule
Method  &  Throughput($\times$Bandwidth)  &  Satisfaction Ratio \\
\midrule
Our Method & 4.61e5 &  95.28\%  \\
\midrule
Equally Dividing &  4.12e5 & 99.02\%  \\
\midrule
Ignoring Demands & 5.18e5 & 91.45\%  \\
\bottomrule
\end{tabular}
\end{center}
\end{table}

\begin{figure}[tb]
    \centering
    \includegraphics[width=0.6\textwidth]{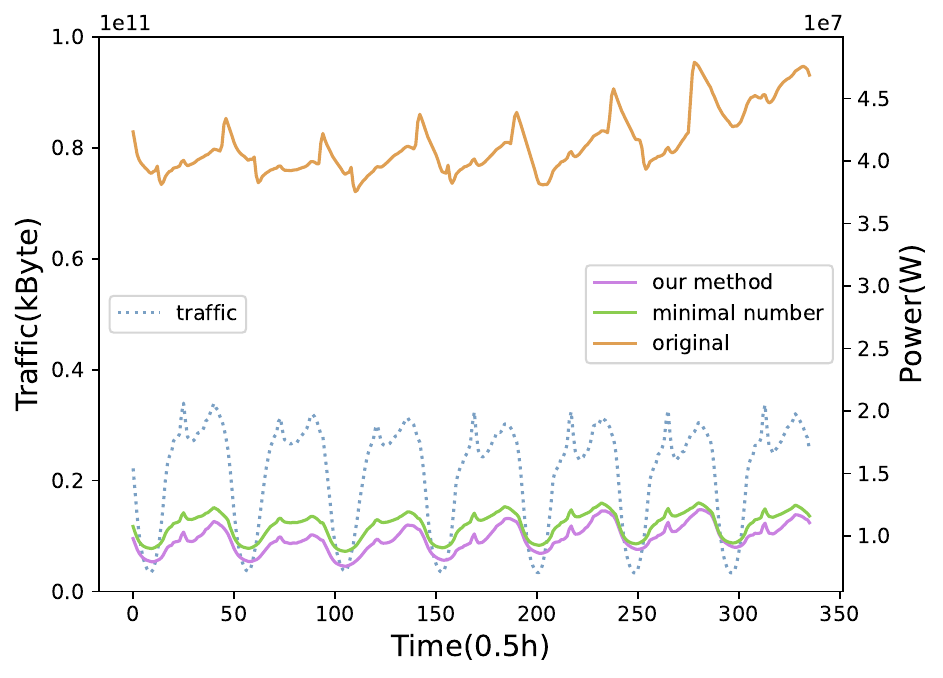}
    \caption{The trend of real traffic and energy consumption under different methods in a week. The dashed line shows the traffic trend, while the solid line displays the change in  energy consumption in a week for our method and two baselines.}
    \label{fig:flow_power}
\end{figure}

\paragraph{Sleep Mechanism}
We evaluate our algorithm using the traffic data from one week in Nanchang. 
In order to confirm the generalization of the algorithm, we only use the data from the first day for training. 
The RL agent should adjust the control statuses of all cells every half an hour to minimize energy consumption and the number of switching.

We compare our method with two baselines.
The former method is original energy consumption, while the latter one~\cite{peng2011traffic} always selects the minimal number of cells in each grid to meet the traffic load.
The experiment result is shown in Fig~\ref{fig:flow_power}.
As shown in Fig.~\ref{fig:flow_power}, the energy consumption of our method is lower than others, and better fits the changing flow trend.

\section{Conclusion}
\label{sec:con}

Next-generation networks' complexity causes modeling and algorithmic deficits in conventional network optimization methods. This article proposes a 6G MNDT architecture. Inner and outer loops make up the proposed MNDT architecture. The inner loop includes an AI-based optimizer and a digital network simulation model. The AI-based optimizer iteratively optimizes based on the simulation engine's feedback. We can overcome modeling and algorithmic deficits with such a simulation-optimization paradigm. After implementing network configuration on a real network, the outer loop updates mirror models and simulation models based on real-world network performance. In practice, we build a network digital twin prototype system to accurately model mobile network elements (e.g., mobile users, base stations), wireless environments, and network performance. An AI-based network optimizer was developed based on the MNDT prototype. The results of the experiments show that the proposed MNDT infrastructure can adapt to the complex environment of mobile networks and provide effective solutions for network optimization.%


\bibliographystyle{unsrt}
\bibliography{Reference}  

\end{document}